\documentstyle[11pt,newpasp,twoside,epsf]{article}
\def\edcomment#1{\iffalse\marginpar{\raggedright\sl#1\/}\else\relax\fi}
\reversemarginpar

\begin{document}
\title{Blazars (VLBA \& GLAST)}
\author{B. Glenn Piner}
\affil{Department of Physics and Astronomy, Whittier College, 13406 E. Philadelphia St.,
Whittier, CA, 90608}

\begin{abstract}
This review paper discusses the past and present observations of
$\gamma$-ray blazars with both space and ground-based $\gamma$-ray telescopes
(such as EGRET and current TeV telescopes), and
with the VLBA.  I also discuss the more sensitive $\gamma$-ray telescopes such as AGILE and GLAST
that will become available over the next several years, 
and how observations
with these telescopes, together with improved VLBA observations,
can increase our understanding of blazar physics.
\end{abstract}

\section{Blazar Overview}
Blazars are those AGN whose apparent luminosity is dominated by beamed emission
from a relativistic jet containing energetic particles (electrons and protons or
electrons and positrons) and magnetic fields, whose bulk 
motion is directed close to our line of sight.
The characteristic blazar spectral energy distribution (SED) is two-peaked, with the low-frequency peak due to synchrotron
radiation, and (in the standard $\gamma$-ray blazar model) the high-frequency peak due to inverse-Compton scattering, either
of the jet's own synchrotron photons (synchrotron self-Compton, or SSC emission), or of external photons
(external Compton scattering, or ECS emission).

Blazars have been claimed to show a continuous change in their spectral energy distribution with luminosity --- the so-called 
``blazar main sequence'' (Fossati et al. 1998; Ghisellini et al. 1998).
The most luminous blazars, flat-spectrum radio quasars, have their
synchrotron peak at mm-IR frequencies and an inverse-Compton peak at $\sim$GeV $\gamma$-ray energies.
The least luminous blazars, high-frequency peaked BL Lac objects, 
have their synchrotron peak at X-ray energies 
and an inverse-Compton peak at $\sim$TeV $\gamma$-ray energies. These differences in peak
frequencies have led to the designations of ``red'' and ``blue'' blazars, respectively. 
The SED differences may arise from increasing radiative losses from ECS emission
in the luminous red blazars, where large external radiation fields are present (Ghisellini et al. 1998).
Following current thinking for the unification of radio-loud sources,
the red blazars and FR II radio galaxies belong to the same population, as do the blue blazars and
FR I radio galaxies.

The differences in the locations of the inverse-Compton peaks between red and blue blazars require that the high-energy
emission from these classes of 
objects be studied by different instruments.  The $\gamma$-ray emission from red blazars must be studied by
space-based pair-production telescopes such as EGRET, while that from the blue blazars can be studied 
by ground-based telescopes or particle detectors sensitive to either the Cerenkov light or energetic 
particles produced in the particle shower caused by a TeV $\gamma$-ray striking the atmosphere.
(Note that the classification of blazars on the blazar sequence as described above, although useful for discussing
the $\gamma$-ray blazars, is almost certainly an oversimplification, as evidenced by the recent finding by
Padovani et al. (2003) of a population of 
flat-spectrum radio quasars with SEDs similar to those of high-energy peaked BL Lacs.)

When imaged with VLBI, blazars display one-sided core-jet structures and many show rapid superluminal motions.
The fastest observed superluminal motions indicate that
bulk Lorentz factors range up to $\Gamma\approx50$ (Jorstad et al. 2001a, with their assumed
cosmological values of $h=0.65$ and $q_{0}=0.1$), and the observed SEDs
indicate particle Lorentz factors up to $\gamma\sim10^{7}$. 
Blazars are thus efficient accelerators, both of bulk plasma and of individual particles.
The question of exactly how these accelerations occur 
makes blazars astrophysically interesting objects. 

The favored model for blazar flares remains the shock-in-jet model.  According to this model, shocks
in the jet caused by interactions of the jet plasma with either density variations in the jet
(the internal shock models, e.g., Spada et al. 2001) or with the external medium
(the external shock models, e.g., Dermer \& Chiang 1998) transfer bulk kinetic energy to internal energy of the plasma,
which then radiates.  Shocks also provide a natural explanation for the ``components'' or ``blobs'' seen
in VLBI images, as has been shown through numerical simulations (e.g., Agudo et al. 2001).

Much current progress in understanding blazar flaring is coming from the results of long-term X-ray monitoring
(e.g., Tanihata et al. 2003).  
The upcoming GLAST $\gamma$-ray mission should have the sensitivity to allow $\gamma$-ray monitoring 
with time resolutions similar to that of these X-ray studies.
Since the high-energy missions and the VLBA both probe the energetic
regions near the blazar central engine, both will be instrumental in answering 
the open questions in blazar physics.

\section{EGRET Blazars}
As one of the four instruments on the {\em Compton Gamma Ray Observatory}, EGRET was a pair-production
telescope that detected 93 blazars (66 with high confidence) and one radio galaxy
during its lifetime from 1991 to 2000 (Hartman et al. 1999). 
EGRET detected large energy outputs in the $\gamma$-ray range, with many sources having an apparent $\gamma$-ray
luminosity greater than that at all other wavelengths.
EGRET also monitored dramatic flaring events; these provided independent confirmation of
relativistic beaming through arguments based on source compactness (e.g., Mattox et al. 1997a).
With the end of the CGRO mission in June 2000, there is currently no active GeV telescope.

The EGRET detections raised significant questions, including why
some strong flat-spectrum radio sources such as 3C~279 were strong EGRET sources
while comparably strong radio sources such as 3C~345 were not detected.
The answer may lie in the different dependences of the radio and 
$\gamma$-ray flux on the jet Doppler factor.
The $\gamma$-ray flux depends on the Doppler factor to a higher power, and so
a higher fraction of high Doppler factor sources is expected in a $\gamma$-ray flux-limited sample,
for SSC and ECS emission.
Lister (1998) and Lister \& Marscher (1999) used Monte Carlo simulations to study
the Doppler factor distributions of radio and $\gamma$-ray selected samples.
They found the $\gamma$-ray selected samples to have on average a higher Doppler factor, and therefore 
a smaller angle to the line-of-sight, and faster apparent speeds.

A number of the EGRET blazars were 
monitored with the VLBA from 1993 to 1997 by Jorstad et al. (2001a; 2001b).
These authors measured a significantly faster distribution of apparent speeds than
those measured for flat-spectrum radio-selected samples, confirming that the $\gamma$-rays
are more highly beamed than the radio emission.
They also found a statistical correlation between $\gamma$-ray flares and the ejection of new
VLBI components.
This correlation
demonstrates a coupling between $\gamma$-ray flares and VLBI components,
consistent with both being different manifestations of the same shock event.

There will have been a number of improvements in both VLBA imaging and in the
interpretation of VLBA images by the time of the launch of GLAST. 
Our understanding of the nature of the components that are
followed on VLBI images, and of their relation to the underlying jet, is
improving through numerical simulations (e.g., Agudo et al. 2001).  
Multi-frequency VLBI polarimetry to measure and apply the necessary Faraday
rotation corrections is now routine (e.g., Zavala \& Taylor 2003).
Full-resolution imaging at 86 GHz is beginning, which will allow imaging inside the current
VLBI core of many of these sources.  Finally,
the results of the VLBA 2 cm survey are available for use as a radio-selected control sample.

\section{TeV Blazars}
Ground-based TeV $\gamma$-ray telescopes have detected a small subset of blue blazars.
The TeV blazars are restricted to relatively nearby sources ($z<0.2$), 
because of the absorption of TeV $\gamma$-rays
by pair-production on the extragalactic background light.
The detection of six blazars in TeV $\gamma$-rays
has been confirmed by multiple detections with different TeV telescopes; all of these are high-frequency peaked BL Lacs.
Four of these confirmed detections were known at the time of this conference:
Markarian~421, Markarian~501,
1ES~1959+650, and H~1426+428 (Horan et al. 2003 and references therein).
Two additional sources have been confirmed as TeV blazars since this conference:
PKS~2155$-$304 (Djannati-Ata\"{\i} et al. 2003), and 1ES~2344+514 (Tluczykont et al. 2003).

We have been 
studying the jets of these TeV blazars with the VLBA for several years.
Earlier observations of Mrk~421 were presented by Piner et al. (1999); more recent
VLBA observations following the prolonged flaring state of Mrk~421 in early 2001
will be presented here and in a future paper.  Observations of Mrk~501 have recently been reported
(Edwards \& Piner 2002; Giroletti et al. 2003).
Observations of 1ES~1959+650,
PKS~2155$-$304, and 1ES~2344+514 are reported by Piner \& Edwards (2003).

All of the TeV blazars have similar jet morphologies and kinematics.
An initial collimated
jet extends to a few mas from the core.  Beyond this region
the jet loses collimation and changes to a diffuse, low-surface-brightness morphology.
The superluminally moving shocks, or components, present in the EGRET blazars
are not apparent in the parsec-scale jets of these sources.  The components in TeV blazar
jets are predominantly stationary or subluminal.
The VLBI cores of these sources are partially resolved and have brightness temperatures
of a few $\times10^{10}$ K (a few $\times10^{11}$ K for the brighter sources Mrk~421 and Mrk~501).

The TeV flaring timescales demand large Lorentz factors in the $\gamma$-ray production region
(from compactness arguments).
The pertinent question about these jets is then: Where are the moving shocks that are presumably responsible for these TeV flares
and that should be visible on the VLBI images?
The most likely explanation for the VLBI properties of these sources seems to be that
the bulk Lorentz factor has been reduced between the TeV emitting scale and the parsec scale,
and that the shocks have dissipated.  This implies a relatively efficient conversion of the bulk
kinetic energy of the jet into particle kinetic energy.
Such a change in the bulk Lorentz factor (from $\Gamma>10$ to $\Gamma$ of a few) also provides an explanation
for problems encountered in BL Lac -- FR I unification (Georganopoulos \& Kazanas 2003).
Note that the jet decollimation observed in all of these sources at a few mas from the core
is also indicative of a jet that has little momentum left and is easily influenced by the
external medium.
Such rapid deceleration may be typical of  high-frequency peaked BL Lacs in general, but in the TeV-detected subset
we have the best evidence for an initially large Lorentz factor.

We have not yet seen any components emerge from the cores
of any of these sources in response to a TeV flare
or prolonged state of increased TeV activity.
The VLBI jets seem decoupled from the activity in the core, in contrast to jets of the EGRET blazars.
Figure~1 shows two VLBA images of Mrk~421 at 22 GHz, taken several months after the period of intense
high-energy activity in early 2001.  Two additional epochs from 2002 are not shown.
No jet components were seen other than those reported by Piner et al. (1999).
However, polarization changes were seen in the nearly stationary components.
The magnetic field of component C6 (the component 1 mas north of the core)
rotates from nearly perpendicular to the jet to
nearly parallel to the jet, while increasing in fractional polarization from 10\% to 30\%,
over the course of the four epochs.
Homan et al. (2002) have reported similar polarization changes in the jet components of other blazars.
If this polarization change represents the motion of some disturbance that originated
with the 2001 $\gamma$-ray flares in the VLBI core, then the apparent speed of that disturbance is about $2 c$.

\begin{figure}[!t]
\plotfiddle{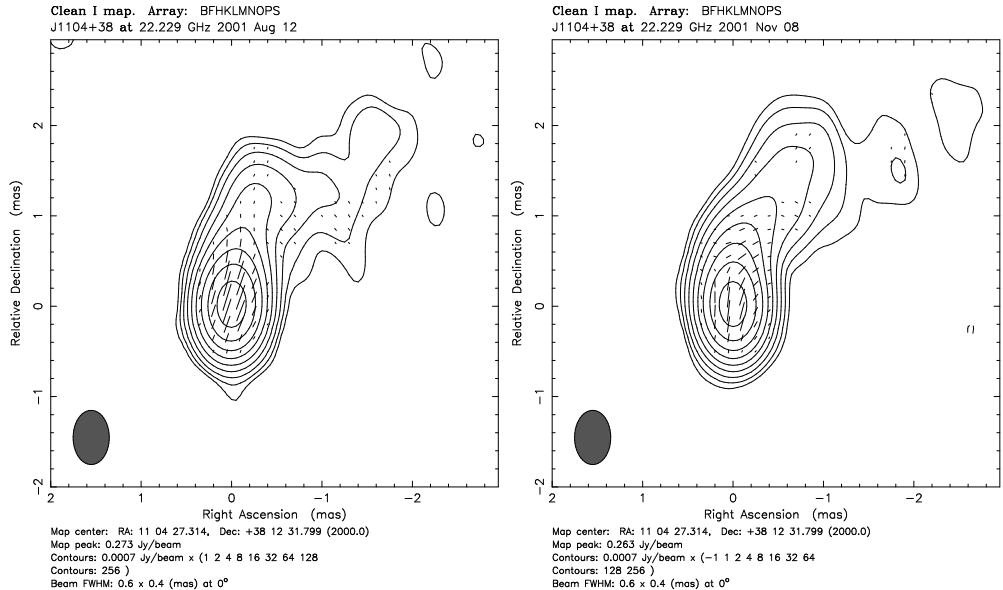}{3.0in}{-90}{50}{50}{-210}{265}
\caption{VLBA Images of Mrk~421 from 22 GHz observations in 2001. The polarization vectors
indicate the EVPA and have a scale of 25 mas/Jy.}
\end{figure}

\section{AGILE and GLAST}
AGILE and the GLAST Large Area Telescope (LAT),
two more sensitive GeV $\gamma$-ray telescopes that will be launched during
the next several years, should detect many more sources than EGRET did --- up to thousands of sources for GLAST.
The relevant characteristics of these two telescopes, in comparison to EGRET, are
summarized in Table~1.  Telescope specifications have been collected from the various
science documents and web pages for each mission\footnote{http://glast.gsfc.nasa.gov/science/,
http://www.roma2.infn.it/research/comm2/agile/, http://www.roma2.infn.it/research/comm2/agile/ai.ps,
http://agile.mi.iasf.cnr.it/Homepage/performances.shtml, http://agile.mi.iasf.cnr.it/Homepage/sources/a-science.ps.gz}.
The numbers quoted for source location apply to relatively bright sources;
the source location is worse for fainter sources.  For example, for sources near the LAT detection threshold,
the diameter of the 95\% confidence region will be $\approx10'$. This may
cause source identification problems (more than one candidate source in the error circle)
for a small number ($\approx$1-2\%) of sources.
The estimates for the number of detected AGN are based on extrapolations of the EGRET luminosity
function by  Stecker \& Salamon (1996), who assumed a linear relation between the radio and
$\gamma$-ray luminosity.
Estimates for the number of AGN detected by GLAST range from $\approx 3000$ to $\approx 11,000$, depending
on the instrument sensitivity assumed.

\begin{table}[!t]
\caption{EGRET, AGILE, and GLAST LAT comparison}
\begin{tabular}{l c c c} \tableline
Property & EGRET & AGILE & GLAST LAT \\ \tableline
Energy Range & 20 MeV -- 30 GeV & 20 Mev -- 50 GeV & 20 MeV -- 300 GeV \\
Field of View & 0.5 sr & 2.5 sr & $> 2$ sr \\
Source Location\tablenotemark{a} & 15$'$ & 7.5$'$ & $<$ 0.5$'$ \\
Sensitivity(cm$^{-2}$ s$^{-1}$)\tablenotemark{b} & $\sim 1\times10^{-7}$ & few $\times10^{-8}$ & few $\times10^{-9}$ \\
Launch & 1991 & 2005 & 2006 \\
\# of AGN detections\tablenotemark{c} & $\sim 100$ & few hundred & few thousand \\ \tableline \tableline
\end{tabular}
\tablenotetext{a}{$1\sigma$ radius, flux $10^{-7}$ cm$^{-2}$ s$^{-1}$ ($>$ 100 MeV), high $|b|$}
\tablenotetext{b}{$>$ 100 MeV, high $|b|$, for exposure of one-year all sky survey, photon
spectral index $-2$}
\tablenotetext{c}{From the $\gamma$-ray luminosity function of Stecker \& Salamon (1996)}
\end{table}

The increased sensitivity of these instruments will improve blazar
observations significantly compared to EGRET results, 
with the most dramatic improvements coming from the $\approx30$ times greater
sensitivity of GLAST.  Dermer \& Dingus (2003) have estimated that GLAST
should detect a bright blazar flare with an integral flux $> 2\times10^{-6}$ photons cm$^{-2}$ s$^{-1}$
(which are the best candidates for Target of Opportunity pointings, temporal and spectral studies,
and coordinated multiwavelength observations) every 3-4 days.
During such flares, GLAST will be able to follow the $\gamma$-ray light curve with improved temporal
and spectral resolution, including detection of spectral index changes during the flare.
These improved $\gamma$-ray measurements, together with coordinated observations at other wavelengths,
should strengthen tests of the dominant radiation mechanisms and $\gamma$-ray emission sites.

\section{GLAST Sources in the Radio}
Since GLAST is likely to detect thousands of sources, the important question for the radio
community is what these faint $\gamma$-ray sources will be like in the radio.
Mattox et al. (1997b) showed that EGRET detections were correlated with flat-spectrum 
radio-loud sources above about 1 Jy at 5 GHz. 
The most significant correlation (99.9994\% confidence) found by Mattox et al. (1997b) was between
the $\gamma$-ray flux and the correlated VLBI flux densities, supporting arguments that the $\gamma$-ray emission
is taking place at the base of a relativistic jet near the compact radio emission.

Since EGRET detections were correlated with sources with compact radio flux densities above about 1 Jy,
we might expect GLAST detections to be correlated with sources with compact radio flux densities above about 30 mJy
--- assuming a 30 times better sensitivity for GLAST and 
using the linear relation between the radio and
$\gamma$-ray luminosity assumed by Stecker \& Salamon (1996).
This opens the possibility that GLAST may detect sources other than blazars,
because radio galaxies, radio-quiet quasars, and Seyfert galaxies can all have compact radio flux densities at the
mJy level.  As with the EGRET detections and 1 Jy radio sources, the GLAST detections should be biased towards
the highest Doppler factor sources from such a radio-selected 30 mJy parent sample (Lister \& Marscher 1999).

We now consider whether there is an appropriate radio catalog from which to draw GLAST candidate sources.
The recently published ``CLASS complete sample'' (Myers et al. 2003) is a complete sample of flat-spectrum
sources ($\alpha>-0.5$ between the 1.4 GHz NVSS and 5 GHz GB6 catalogs) over 30 mJy (in the GB6 catalog).
This sample consists of $\approx12000$ sources between $0\deg<\delta<75\deg$, excluding the galactic plane.
The published survey presents VLA A configuration 8.4 GHz images of $\approx11000$ of these sources.
The CLASS complete sample can be considered as a rough GLAST candidate source list
for the northern hemisphere.

An even better correlation should be obtained between the VLBI flux densities of sources in the CLASS
sample and GLAST detections, if VLBI flux densities are measured.
While the CLASS complete sample will be quite useful, it is likely to be missing a subset of GLAST sources.
Objects with enough extended flux to be steep spectrum
($\alpha<-0.5$) in single-dish or VLA surveys may still have compact flat-spectrum cores above the 30 mJy
level, and these cores may be detected by GLAST (for example, the one radio galaxy detected by EGRET, Centaurus A,
would not be in the CLASS complete sample, even if it met the declination criteria).
Some effort will be needed to identify these candidate GLAST steep-spectrum sources.
With the likely detection of thousands of new $\gamma$-ray sources in the next several years, the
characterization of candidate source properties at other wavebands  
(including redshift measurements at optical wavelengths) will be an essential
step toward understanding the physics of the $\gamma$-ray emission in these sources.

\acknowledgements
Seth Digel and Jim Ulvestad
provided information and assistance for this talk.

\end{document}